\documentclass[prb,10pt,twocolumn,showpacs,superscriptaddress]{revtex4}
\usepackage[dvips]{epsfig}
\usepackage{float}
\usepackage{amsfonts}
\usepackage{amsmath}
\usepackage{amssymb}
\usepackage{enumerate}

\usepackage{hhline}
\usepackage{threeparttable}
\usepackage{supertabular}
\usepackage{multirow}
\usepackage{tabularx}
\usepackage{pslatex}
\usepackage{setspace}

\newcolumntype{Y}{>{\centering\arraybackslash}X}

\begin{document}

\title{Photoluminescence measurements of quantum-dot-containing
semiconductor microdisk resonators using optical fiber taper waveguides}

\author{Kartik Srinivasan}
\email{kartik@caltech.edu}
\affiliation{Thomas J. Watson, Sr., Laboratory of Applied Physics, California Institute of Technology, Pasadena, California 91125}
\author{Andreas Stintz}
\affiliation{Center for High Technology Materials, University of
New Mexico, Albuquerque, New Mexico}
\author{Sanjay Krishna}
\affiliation{Center for High Technology Materials, University of
New Mexico, Albuquerque, New Mexico}
\author{Oskar Painter}
\affiliation{Thomas J. Watson, Sr., Laboratory of Applied Physics, California Institute of Technology, Pasadena, California 91125}
\date{\today}

\begin{abstract}
Fiber taper waveguides are used to improve the efficiency of room temperature
photoluminescence measurements of AlGaAs microdisk resonant cavities with embedded self-assembled InAs quantum dots.  As a
near-field collection optic, the fiber taper improves the
collection efficiency from microdisk lasers by a factor of
$\sim$ 15-100 times in comparison to conventional normal incidence free-space collection techniques. In addition, the fiber taper can serve as a
efficient means for pumping these devices, and initial
measurements employing fiber pumping and collection
are presented. Implications of this work towards chip-based cavity quantum 
electrodynamics experiments are discussed.
\end{abstract}
\maketitle

\setcounter{page}{1}
\section{Introduction}
\label{sec:intro}

The ability to efficiently couple light into and out of
semiconductor microcavities is an important aspect of many microphotonic technologies\cite{ref:Yamamoto2}, and plays a vital role in chip-based implementations of cavity quantum electrodynamics (cQED) for quantum networking and cryptography\cite{ref:Mabuchi,ref:Kimble2,ref:Barnes2}.  While some geometries, such as micropillar cavities, exhibit highly directional 
emission that can be effectively collected\cite{ref:Solomon,ref:Gerard1}, 
coupling to wavelength-scale semiconductor microcavities is in general 
non-trivial\cite{ref:Slusher1,ref:Levi2,ref:Nockel,ref:Barnes2}, due to a number of factors.  These
include the size disparity between the modes of the microcavity
and those of standard free-space and fiber optics, as well as the
refractive index difference between semiconductors and glass or
air, as well as the potentially complicated cavity mode profiles
sustained by these devices.  One technique that we have recently employed to couple efficiently to semiconductor microcavities is evanescent
coupling through an optical fiber taper
waveguide\cite{ref:Srinivasan7}.

The fiber taper\cite{ref:Birks,ref:Birks_OFC} is simply a standard single mode fiber
that has been heated and stretched down to a minimum diameter on
the order of a wavelength.  Such fiber tapers have been used as near-ideal coupling channels for glass-to-glass coupling with silica-based microcavities such as
microspheres\cite{ref:Knight,ref:Dubreuil1,ref:Cai,ref:Spillane2} and
microtoroids\cite{ref:Armani}.  Our recent experiments have
indicated that they can also serve as efficient couplers to high-refractive index semiconductor-based devices, such as photonic crystal
waveguides\cite{ref:Barclay5}, photonic crystal
cavities\cite{ref:Srinivasan7,ref:Barclay7} and
microdisks\cite{ref:Borselli,ref:Srinivasan9}.  Here, we consider
the use of fiber tapers within active semiconductor devices consisting of AlGaAs microdisk cavities
with embedded InAs quantum dots\cite{ref:Srinivasan9}.

In Section \ref{sec:introduction}, we qualitatively describe the
issues addressed in this paper, as well as the methods used in
device fabrication and the experimental setup we use. In Section
\ref{sec:passive_taper}, we briefly review passive measurements in the $1200$ nm wavelength band to determine the intrinsic optical losses of the optical resonant cavities under study.  In Section \ref{sec:fs_pump}, we
present experimental results demonstrating the improvements that
result when free-space collection is replaced by fiber-based
collection in photoluminescence measurements, while in Section
\ref{sec:fiber_pump}, we present initial results on microdisk
lasers that employ both fiber pumping and fiber collection.
Finally, in Section \ref{sec:discussion}, we consider some of the
applications of this work to future experiments.

\section{Preliminary discussion and experimental methods}
\label{sec:introduction}

\begin{figure}[ht]
\begin{center}
\epsfig{figure=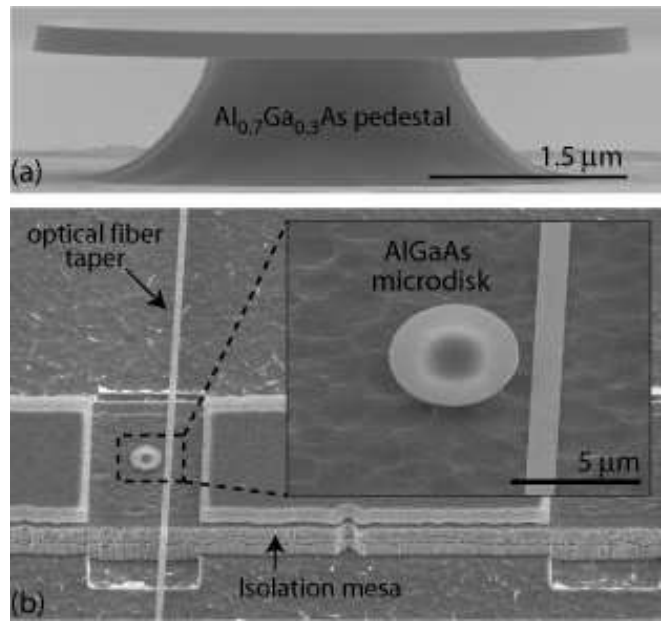, width=\linewidth}
\caption{(a)-(b) Scanning electron microscope (SEM) images of
fabricated microdisk structures. Image (b) shows the isolation mesa that is incorporated in order to
aid in the taper testing.  An optical fiber taper
aligned to the side of a microdisk is also visible in this image.}
\label{fig:udisk_SEMs}
\end{center}
\end{figure}

The specific devices we consider in this work are AlGaAs microdisk
cavities with embedded quantum dots (QDs).  The epitaxy
used is shown in Table \ref{table:DWELL_epi}, and
consists of a single layer of InAs quantum dots embedded in an
InGaAs quantum well\cite{ref:Liu_G}, which is in turn sandwiched between layers of
AlGaAs and GaAs to create a 255 nm thick waveguide.  This DWELL
(short for dot-in-a-well) material has a room temperature ground
state emission peak at around 1190 nm (Fig.
\ref{fig:passive_meas}(a)), and is grown on top of a 1.5 $\mu$m
Al$_{0.70}$Ga$_{0.30}$As layer that eventually serves as a support pedestal for the microdisk.  Fabrication of the microdisks (Fig.
\ref{fig:udisk_SEMs}) is accomplished through the following series
of steps: (i) deposition of a 200 nm Si$_x$N$_y$ mask layer, (ii)
electron-beam lithography and a subsequent reflow of the resist,
(iii) inductively-coupled plasma reactive ion etch (ICP-RIE) of
the Si$_x$N$_y$ layer, (iv) ICP-RIE of the QD-containing waveguide
layer, (v) photolithography and isolation of the microdisk onto a
mesa stripe that is several microns above the rest of the chip
(Fig. \ref{fig:udisk_SEMs}(b)), and (vi) HF acid etch of the
Al$_{0.70}$Ga$_{0.30}$As layer to form the pedestal which supports
the disk.  The fabricated microdisks in this work are $D\sim$4.5 $\mu$m in diameter.  

\renewcommand{\arraystretch}{1.2}
\begin{table}
\caption{Epitaxy for 1-DWELL microcavity lasers.}
\label{table:DWELL_epi}
\begin{center}
\begin{tabularx}{\linewidth}{YYY}
\hline
\hline
Layer & Materials & Thickness \\
\hline
Surface cap layer & GaAs & 100 $\overset{\circ}{\text{A}}$ \\
Top waveguide layer & Al$_{0.30}$Ga$_{0.70}$As & 400 $\overset{\circ}{\text{A}}$\\
Top waveguide layer & GaAs &  740 $\overset{\circ}{\text{A}}$ \\
Quantum well layer & In$_{0.15}$Ga$_{0.85}$As &  60 $\overset{\circ}{\text{A}}$ \\
Quantum dot layer & InAs & 2.4 monolayer \\
Barrier layer     &  In$_{0.15}$Ga$_{0.85}$As &  10 $\overset{\circ}{\text{A}}$\\
Bottom waveguide layer & GaAs & 740 $\overset{\circ}{\text{A}}$ \\
Bottom waveguide layer & Al$_{0.30}$Ga$_{0.70}$As & 500 $\overset{\circ}{\text{A}}$ \\
Sacrificial buffer layer & Al$_{0.70}$Ga$_{0.30}$As & 15000 $\overset{\circ}{\text{A}}$ \\
Substrate & GaAs & N/A \\
\hline
\hline
\end{tabularx}
\end{center}
\end{table}

The free-space collection from a whispering gallery mode (WGM) of
a microdisk is a function of a number of factors, including the
position and numerical aperture (NA) of the collection lens, the radiation pattern, and the quality factor ($Q$) of the resonant mode.  Optical losses from the microdisk include not only the (ideal) radiation due to radial tunneling of light from the disk periphery, but also scattering losses due to surface roughness imperfections at the disk edge and material absorption.  For high-refractive index ($n \sim 3.5$) III-V semiconductor microdisks, surface roughness scattering is typically the dominate form of radiation from the microcavity.  The intrinsic radiation loss of semiconductor microdisks is almost negligible in all cases, save the smallest of microdisks; the radiation $Q$ of the lowest radial order WGM of the $D \sim 4.5$ $\mu$m microdisks studied here is greater than $10^{16}$ at the QD emission wavelength of $1200$ nm, and is greater than $10^6$ for $D \sim 1.5$ $\mu$m.  As such, any light that is collected through free-space methods is the result of scattering of the WGM off 
imperfections in the microdisk\cite{ref:Levi2}, a relatively
inefficient and non-directional process. Bulk material absorption and absorption due to surface states also play a role, and for even high-purity (nominally undoped) AlGaAs microdisks we have found absorption losses on par with scattering losses in the $1.2$-$1.5$ $\mu$m wavelength range for $D \sim 5$ $\mu$m microdisks fabricated using the above described procedure (note that absorption losses rapidly increased below $1$ $\mu$m wavelength, an effect we are currently studying further).  This results in a situation where the more perfect the microdisk is made (through reduction in surface roughness), and the further the $Q$ factor is improved, the more difficult it becomes to collect light from the resonant modes.  Although there may be some potential in
modifying the disk geometry\cite{ref:Levi2,ref:Barnes2} to improve this situation (for
example, by etching a shallow second-order grating in the microdisk surface), the ability
to do this while maintaining high $Q$ factors could be 
of potential difficulty.  The most successful method to date for increasing collection efficiency from semiconductor microdisk resonators seems to be placement of the collection optics in the plane of the disk\cite{ref:Gayral,ref:Michler2}, resulting in more effective capture of the predominantly low-angle scattered light. 

The fiber taper offers an attractive alternative because it
provides a means to directly couple light out of the
WGMs, without relying upon the weak intrinsic radiation of the microdisk or the non-directionality of surface roughness scattering.  This evanescent near-field coupling, which is a function of the integrated modal
overlap of the microdisk and taper modes over the interaction
region\cite{ref:Manolatou,ref:Borselli_coupling}, has been
demonstrated to be appreciable in previous
works with small diameter semiconductor microdisks\cite{ref:Borselli,ref:Srinivasan9} where phase-matching between the glass fiber taper waveguide and the semiconductor microdisk is not as limiting.  While the fiber taper
does load the cavity mode, and thus degrade its $Q$, the key point is
that the added loss is primarily \emph{good loss} in the sense that it
can be efficiently collected into the taper mode of
interest\cite{ref:Spillane2,ref:Barclay7}.  This allows for the loaded $Q$ to be maintained at a high value while simultaneously obtaining high collection efficiency\cite{ref:Borselli,ref:Srinivasan9}.  The situation is analogous to the
case of a Fabry-Perot cavity where one mirror is intentionally
made to have a slightly lower reflectivity for output coupling,
which limits the $Q$ of the cavity, but not beyond some acceptable level. While in that case, the
cavity $Q$ is fixed by the mirror reflectivities, here we have
some flexibility over the $Q$ and the amount of loading by
adjusting the cavity-taper separation.

\begin{figure*}[ht]
%\begin{center}
\epsfig{figure=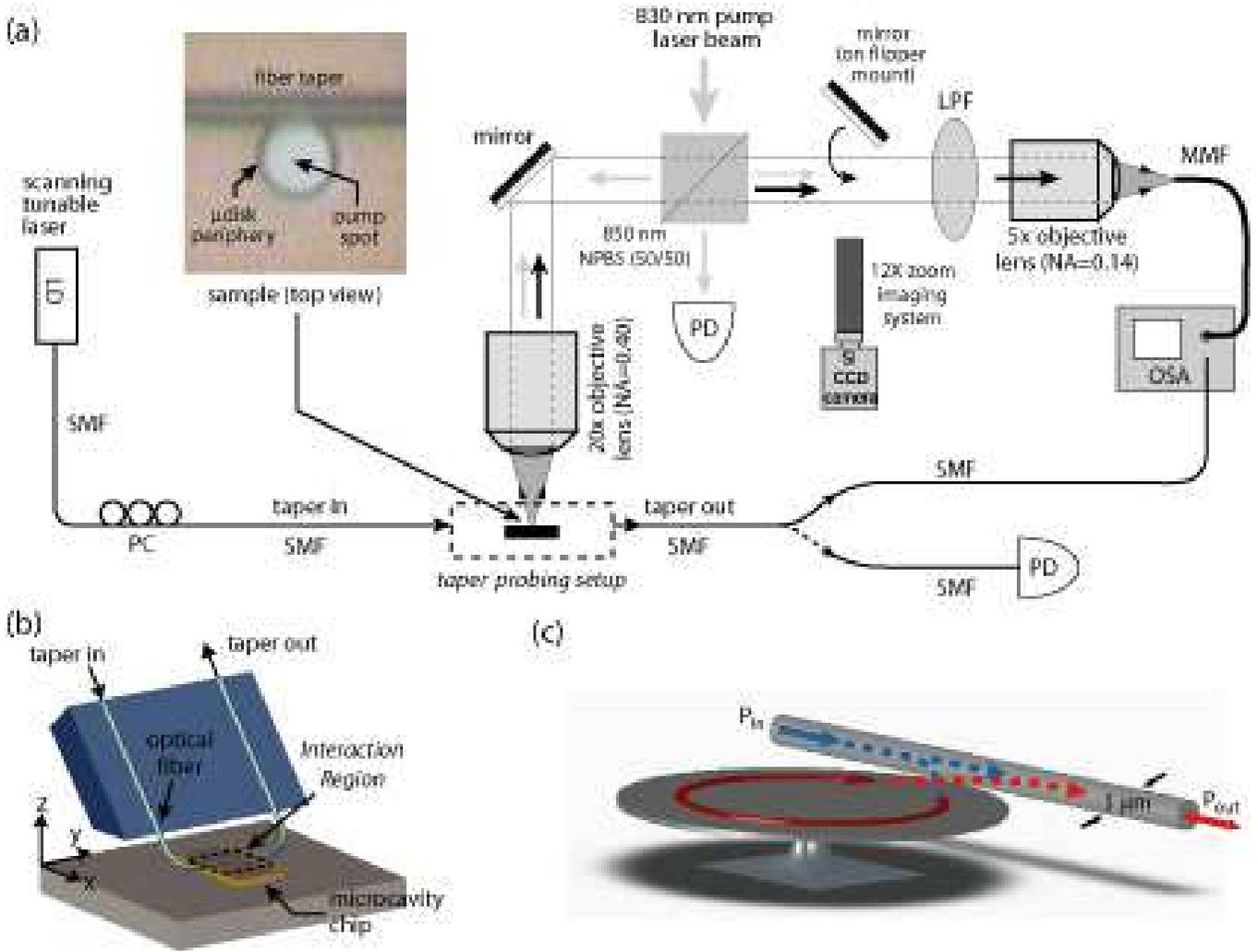, width=\linewidth}
%\begin{spacing}{1.2}
\caption{Note: Figure is to extend over both columns. (a)Experimental setup for studying the QD-microdisk devices. The sample is mounted on an X-Y stage with 50 nm encoded resolution for positioning, and a pump beam at $\lambda_p=830$ nm is directed through an ultra-long working distance objective lens (NA = 0.40) at normal incidence to the sample surface.  The free-space pump laser power is monitored by using a 830 nm wavelength 50/50 non-polarizing beamsplitter (NPBS) with a calibrated photodetector (PD) on one of the ports.  The QD free-space photoluminescence in the 1200 nm band is collected at normal incidence from the sample surface using the same objective lens for pump focusing, is transmitted through the 830 nm NPBS and a long-pass pump rejection filter (LPF), and is finally collected into a multi-mode fiber (MMF) using an objective lens with NA = 0.14.  The pump laser and photoluminescence beams are shown as light gray and black arrows, respectively.  To allow for fiber taper measurements, the fiber taper is strung across the sample and positioned in the near-field of the microdisk from above, thus allowing simultaneous (normal incidence) free-space and fiber taper optical pumping and photoluminescence collection.  The output of the fiber taper can either be connected to an InGaAs photodetector (PD) for wavelength scans using the swept tunable laser source in the 1200 nm band, or to an optical spectrum analyzer (OSA) for analysis of the photoluminescence from the microdisk.  The OSA is also used to analyze the free-space photoluminescence using a MMF input. Alignment of the pump beam and the fiber taper to the microdisk is performed by imaging through the pump and collection objective lens, as shown in the inset (a mirror flips in-and-out of the free-space photoluminescence beam path to direct the image to a 12X zoom imaging system).  (b) Schematic of the fiber taper probing geometry in which the fiber taper is mounted in a ``u''-shape configuration on a Z-axis stage with 50 nm encoded resolution for near-field positioning of the taper. (c) Schematic of the taper-to-microdisk interaction region, showing the resonant fiber taper coupling to WGMs of the microdisk.} 
\label{fig:setup}
%\end{spacing}
%\end{center}
\end{figure*}

To compare free-space and fiber-taper-based collection, we use the
experimental setup depicted in Fig. \ref{fig:setup}, which
consists of a fiber taper probing station that has been
incorporated into a standard photoluminescence (PL) measurement
setup.  The taper probing station consists of a motorized X-Y stage (50 nm encoded resolution) on which the
microcavity chip is placed, while the fiber taper waveguide is held in a
``u''-shaped configuration on an acrylic mount as shown in Fig. \ref{fig:setup}(b).  The acrylic fiber mount is attached to a separate motorized Z-axis stage (50 nm
encoded resolution) so that the fiber taper can be precisely aligned to the microdisk (the taper moves vertically and the sample moves in-plane).  The entire taper probing setup (motorized stages, microcavity
chip, and fiber taper waveguide) is mounted onto a larger manually actuated X-Y-Z stage that is positioned underneath
a ultra-long working distance objective lens (NA = 0.4).  This microscope objective is
part of a PL setup that provides normal incidence pumping
and free-space collection from the samples.  The pump laser in the majority of the measurements is a 830 nm laser diode that
is operated in quasi-continuous-wave operation (280 ns pulse
width, 300 ns period).  The pump beam is shaped into a 
Gaussian-like profile by sending the laser beam through a section of single mode optical
fiber, and is then focused onto the sample with a spot size that is
slightly larger than the size of the microdisk
(area $\sim 18$ $\mu$m$^2$). Luminescence from the microdisks
is wavelength resolved by a Hewlett Packard 70452B optical spectrum analyzer.  All of the measurements presented here were performed with the sample maintained in a room temperature environment, with no active cooling or temperature control.

This integrated setup allows for a number of different
measurements to be made.  Passive measurements of the microdisk resonant modes are performed by connecting the input of the fiber taper to a 1200 nm band scanning
tunable laser.  The polarization of the tunable laser output at the fiber taper interaction region with the microdisk (coupling to the WGMs of the microdisk is polarization sensitive) is controlled using a paddle wheel polarization controller (PC).  The light transmitted past the microdisk resonator is fed from the optical fiber to a photodetector in order to monitor the wavelength-dependent
transmission.  Photoluminescence measurements can be done in any
of four potential configurations (i) free space pumping, free
space collection: here, the fiber taper plays no role, and the
vertically emitted power from the disks is collected into a multimode optical fiber
that is then fed into the OSA; (ii) free space pumping, fiber taper 
collection: here, the output of the fiber taper is connected to the OSA; (iii) fiber taper
pumping, free space collection: here, the input of the fiber taper
is connected to a fiber-coupled pump laser; (iv) fiber taper pumping,
fiber taper collection: here, the free-space optics used in the standard PL measurements play no role.

\section{Measurement of cavity $Q$ in the $1200$ nm wavelength band}
\label{sec:passive_taper}

\begin{figure}[ht]
\begin{center}
\epsfig{figure=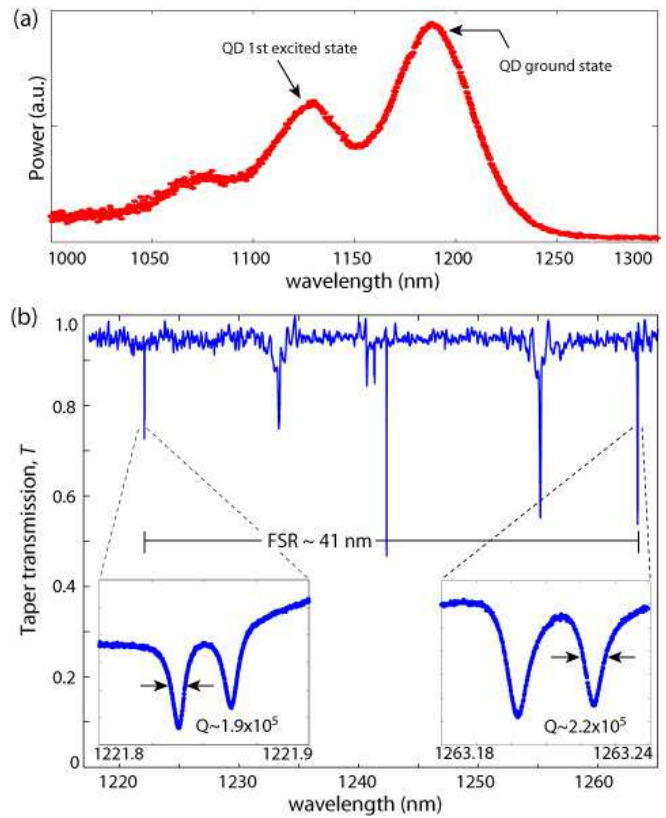, width=\linewidth}
\caption{(a) Photoluminescence from an unprocessed region of the 1DWELL
material. (b) Passive taper-based measurements of a microdisk
in the 1200 nm band.   The insets show high resolution
scans for two sets of doublet modes for this device.  These 
high resolution scans are taken when the taper-microdisk separation 
is a few hundred nm and the depth of coupling is $\sim$5-10 $\%$.} 
\label{fig:passive_meas}
\end{center}
\end{figure}

The devices studied in this work have been previously
characterized in the 1400 nm band, and $Q$s as high as
$3.6{\times}10^5$ have been measured\cite{ref:Srinivasan9}.
Those measurements were done at wavelengths significantly
red-detuned from the QD emission band, where QD absorption and
material absorption in the GaAs/AlGaAs waveguide layers are
expected to be quite small.  To confirm that the cavity $Q$s are
still high near the ground-state QD emission wavelength (peaked near $1190$ nm as shown in Fig. \ref{fig:passive_meas}(a)), we perform passive fiber-taper-based measurements\cite{ref:Srinivasan7,ref:Borselli,ref:Srinivasan9} in the $1200$ nm
band\footnote{The WGMs in the $1400$ nm wavelength band are expected to have very similar radiation and scattering losses as those in the $1200$ nm band for the microdisk geometries studied here.  Differences in $Q$ at these two wavelengths are thus expected to be
indicative of wavelength-dependent material absorption losses.}.
The input of the taper is connected to a fiber-coupled scanning tunable laser
with a wavelength range of $1215$-$1265$ nm and a spectral resolution
of better than $0.01$ pm.  The transmission past the microdisk is monitored by connecting the taper output to a fiber-pigtailed photodetector.  The taper-WGM coupling is sensitive to polarization, and here coupling was optimized for TE-like WGMs with electric field polarized predominantly in the plane
of the microdisk.  The fiber taper is brought into the near-field of the microdisk by first vertically aligning the fiber taper within the plane of the microdisk and then bringing it in from the side towards the disk edge (Fig. \ref{fig:setup}(c)).  Resonances begin to appear within
the taper's transmission spectrum when it is several hundred nanometers to the side
of the microdisk.  The high-$Q$ resonances within the transmission spectrum, under closer inspection, are seen to consist of a pair resonances dips (inset to Fig. \ref{fig:passive_meas}(b)).  The two resonance dips (\emph{doublets}) correspond to standing wave modes that are formed when
surface scattering couples and splits the initially degenerate
clockwise and counterclockwise traveling wave WGMs of the microdisk.  Such doublet resonances have been observed by a number of authors for different whispering-gallery geometries and materials\cite{ref:Weiss,ref:Kippenberg,ref:Borselli}.

Finite-element frequency mode solutions\cite{ref:Spillane3} of the WGM resonances of the microdisks studied in this work ($D = 4.5$ $\mu$m) show that the free spectral range (FSR) is $\sim 40$ nm in the $1200$ nm wavelength band for TE-polarized
modes of low radial mode number ($q=1,2,3,4$).  Higher radial order WGMs ($q \ge 5$) are expected to show up only very weakly in the fiber taper transmission owing to their small radiation limited Q factors ($\lesssim 10^4$) and significantly larger overlap with the support pedestal.  From the broad spectral wavelength scan shown in Fig. 
\ref{fig:passive_meas}(b), a pair of deeply coupled resonant modes separated by a full FSR are observed ($\lambda \sim 1222$ and $1263$ nm), as well as several other deeply coupled resonant modes.  Due to the extended nature of the higher order radial modes and their better phase-matching to a low-index glass waveguide such as the fiber taper, the coupling to the lowest order $q=1$ WGM is typically lower than that of the $q=2$ mode for similar sized microdisks\cite{ref:Borselli,ref:Borselli_coupling}.  We believe that these doublet modes at $\lambda \sim 1222$ nm and $\lambda \sim 1263$ nm are first order ($q=1$) radial modes, while the mode at $\lambda \sim 1242$ nm is probably
a $q=2$ radial mode.  The broader and more weakly coupled intermediate modes are most likely higher
order radial modes, $q=3,4$ (higher order slab modes in the vertical direction of microdisk are also a possibility, though less 
likely due to their reduced radiation $Q$).  Examining the linewidth of the doublet resonances when the taper is 
relatively far away from the microdisk gives an estimate for the cold-cavity, unloaded $Q$ of the modes.  $Q$s as high as $2.2{\times}10^5$ at $\sim$1260 nm and as high as $1.9{\times}10^5$ at $\sim$1220 nm are measured in these microdisks
(insets of Fig. \ref{fig:passive_meas}(b)), the latter of which is only 30 nm red-detuned from the 
peak of the QD emission spectrum (Fig. \ref{fig:passive_meas}(a)). These $Q$ factors 
are still quite high for a wavelength-scale AlGaAs
microcavity\cite{ref:Gayral,ref:Michler2,ref:Yoshie3,ref:Loffler,ref:Badolato},
and correspond to a cavity decay rate of $\kappa/2\pi\sim0.6$ GHz for resonant modes with an effective mode volume of only $V_{\text{eff}} \sim 6(\lambda/n)^3$.  Nevertheless, some degradation in the quality factors from those previously measured in the $1400$ nm band are observed.  These are
believed to be at least in part due to absorption in the QD
layers, as evidenced by the emission in the PL spectrum at these
wavelengths (Fig. \ref{fig:passive_meas}(a)).
%Additional measurements in the $980$ and $850$ nm wavelength bands on similar microdisk resonators formed in AlGaAs indicate a trend of optical loss which significantly increases below a wavelength of $1$ $\mu$m.  This trend in optical loss is similar to that reported in Ref. \cite{ref:Karkhanehchi}, where material absorption that extended $350$ meV within the bandgap was attributed to incorporation of oxygen impurities into the AlGaAs lattice.

\section{Improved collection efficiency with fiber tapers}
\label{sec:fs_pump}

\begin{figure}[ht]
\begin{center}
\epsfig{figure=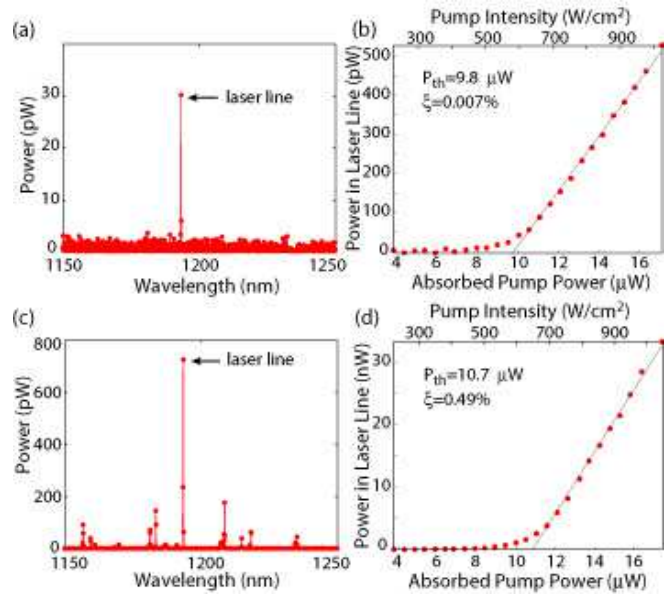, width=\linewidth}
\caption{(a) Emission spectrum and (b) Light-in-vs-Light-out (L-L) curve for normal incidence free-space collection from a (free-space) optically pumped microdisk. Fiber taper collected (c) emission spectrum and (d) L-L curve from the same microdisk resonator, under identical pumping conditions.  The emission spectrums for both the free-space and fiber taper collection configurations were taken near laser threshold at $\sim 580$ W/cm$^2$ of incident pump beam intensity.  The fiber taper collected power included that from the forward propagating transmission channel only.  Unless otherwise specified,
in all of the measurements described in this work, a detector
sensitivity of $\sim 1$ pW is used, and the resolution bandwidth of
the OSA is set to $0.1$ nm.} \label{fig:fs_pump_fs_vs_fiber_coll}
\end{center}
\end{figure}

We now turn to the heart of the current work, which is a study of
the gains in efficiency that can be achieved by using optical\cite{ref:Borselli,ref:Srinivasan9}
fiber tapers as a collection tool in PL measurements.  This is
initially done by comparing the amount of power obtained in
free-space and fiber taper collection configurations, while maintaining 
identical free-space pumping conditions (in terms of pump-beam
intensity and pump beam position).  The free-space collection for
a microdisk that has been pumped at normal incidence with
$\sim580$ W/cm$^2$ at 830 nm is shown in Fig.
\ref{fig:fs_pump_fs_vs_fiber_coll}(a).  This pump intensity incident is near the laser threshold 
for this device (see below), and we see that the peak height at 
$\lambda \sim 1193.5$ nm is $\sim 30$ pW.  For comparison to the fiber taper collection described below, an estimate of the optical losses in the free-space collection setup were made (after removal of the pump rejection filter).  By measuring the collected pump laser power reflected off of the mirror-quality surface of the AlGaAs epitaxy, and assuming a $30\%$ reflection coefficient from the AlGaAs surface, $43\%$ of the reflected pump beam was collected into the OSA.  Additional limitations in the normal incidence free-space collection stem from the finite numerical aperture of the collecting lens (NA=0.4) which covers only $4\%$ of the full $4\pi$ steradians.

Next, we consider the use of the optical fiber taper as a
collection optic in the PL measurements. To obtain an estimate of the 
amount of coupling between the taper and the microdisk, the free-space pump 
beam is blocked, and passive measurements at wavelengths that are slightly red-detuned from the
QD emission are performed as described above in Section
\ref{sec:passive_taper}. Since the FSR for the low radial number WGMs of the microdisks studied here is $\sim41$ nm in the 1200 nm wavelength band, the modes coupled to and studied passively are typically a single FSR red-detuned 
from the lasing mode.  For most experiments, the taper is placed in direct 
contact with the top edge of the microdisk, which increases the amount of coupling from that shown in Fig. \ref{fig:passive_meas}(b) to transmission depths between $30-60\%$.  For this initial measurement, a resonance depth of $\sim$38$\%$ is obtained for a
cavity mode at $\lambda \sim 1238.1$nm, which gives us a
qualitative estimate for the coupling to WGMs overlapping the peak of the gain spectrum\footnote{The coupling between
the taper and microdisk can be different for different cavity modes, 
so this technique is used primarily as a qualitative guide.}. 
This coupling depth corresponds to a taper collection efficiency $\eta_{0} \sim 11\%$, where $\eta_{0}$ is defined\cite{ref:Barclay7}
as the fraction of the optical power from the cavity resonant mode that is coupled into
the fundamental fiber taper mode in the forward propagating transmission direction.  Other loss channels from the microdisk include intrinsic loss of
the cavity in absence of the taper, parasitic coupling into higher-order, non-collected
modes of the fiber taper, and for the standing wave modes studied here, coupling into the backwards propagating fundamental taper mode.  For the moderate coupling depths measured here, the taper coupling efficiency into the forward and backward propagation directions is approximately equal, thus yielding an overall taper coupling efficiency of $\eta_{0}^\prime \sim 22\%$ for this WGM.

Once this level of coupling has been achieved, the tunable laser
output is blocked, the free-space pump beam is unblocked, and the
output of the fiber taper is disconnected from the photodetector
and connected to the OSA to measure the emitted power from the
microdisk.  Fig. \ref{fig:fs_pump_fs_vs_fiber_coll}(c) shows the resulting
spectrum collected by the fiber taper in the forward propagating transmission direction.  We see that at
the wavelength $\lambda \sim 1193.5$ nm, the peak height is $\sim 725$ pW, 
which is nearly a factor of 25 times improvement over
the peak height observed in normal incidence free-space collection.  In addition, a
significant amount of power is present within modes that were not detectable in the free-space case (the noise floor of the OSA was approximately $1$ pW), due to the poor efficiency
of collection in this configuration.

This straightforward comparison of the collected powers for a
single pump power is not necessarily the most appropriate
comparison, however.  The reason for this is that the fiber taper
loads the cavity, thus decreasing the $Q$ of the resonant modes and increasing the
threshold pump power, so that for a given pump power the laser is not equally above threshold in the two measurements. Another, more appropriate comparison is the differential
collection efficiency above threshold, which we label $\xi$. This
is determined by measuring a light-in-light-out (L-L) curve for
the microdisk and taking the slope of this curve above threshold. 
In these curves, the light-out is taken to be the total power within
the laser line, while the light-in is taken to be the estimated absorbed pump power.
The absorbed pump power is determined by multiplying the pump beam intensity by the area of
the microdisk to get an incident pump power (the beam overlaps the 
entirety of the disk), and then multiplying this value by the
absorption of the microdisk at 830 nm.  We estimate this
absorption to be $\sim10\%$, assuming an absorption coefficient of
10$^4$ cm$^{-1}$ in the GaAs, quantum well, and QD
layers\cite{ref:Coldren}, and a reflection coefficient of $30\%$
at the GaAs-air interfaces at the top and bottom of the disk. The
resulting L-L curves are shown in Fig.
\ref{fig:fs_pump_fs_vs_fiber_coll}(b),(d) for both free-space and
fiber-taper collection. We see that the threshold pump power has
indeed increased in the case of fiber-taper collection, but that
the differential efficiency has also significantly improved and is
more than 70 times that of the free-space value.

\begin{figure}[ht]
\begin{center}
\epsfig{figure=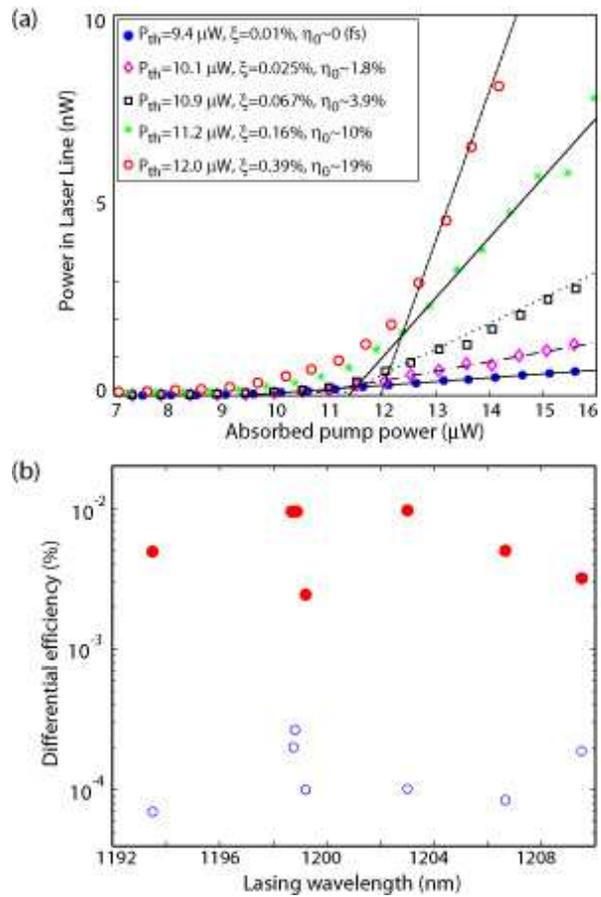,
width=\linewidth} \caption{(a) L-L curves for free-space
pumping and fiber taper collection at different taper positions.  For
each curve, we note the threshold pump power
($P_{\text{th}}$), the fiber taper collection efficiency ($\eta_{0}$) for a mode that is
red-detuned from the peak QD emission, and the above threshold differential efficiency ($\xi$).
(b) Scatter plot of the differential efficiency for fiber taper (filled circles) and
free-space collection (open circles) for a number of different microdisk lasers.  For these measurements the fiber taper collected power included that from the forward propagating transmission channel only.}
\label{fig:fs_pump_fiber_coll_many_L_L}
\end{center}
\end{figure}

To study the tradeoffs between $\xi$ and threshold more
closely, in Fig. \ref{fig:fs_pump_fiber_coll_many_L_L}(a) several
L-L curves are plotted, each for a different taper position with respect to the microdisk (note that the microdisk studied here is not the same as
the one studied above, but the qualitative behavior is identical).
The different taper positions correspond to a varying level of
coupling between the microdisk and taper, which we again qualitatively
estimate through measurements of the coupling to a microdisk WGM  
that is red-detuned from the QD emission in the 1200 nm band. From
Fig. \ref{fig:fs_pump_fiber_coll_many_L_L}(a), we see that in
general, both the threshold power and $\xi$ increase with increasing $\eta_{0}$.  As might be expected, in the course of these measurements it was possible in some cases to load the 
microdisk strongly enough to degrade the initial laser mode's $Q$ to the point that it 
no longer lases, and a different mode (with a higher loaded $Q$) begins to lase.

A number of different microdisk devices have been studied, and the
results described above are fairly consistent from device to device, with $\xi$ routinely 1-2 orders
of magnitude larger when fiber taper collection is employed.  A
scatter plot for some of this data is shown in Fig.
\ref{fig:fs_pump_fiber_coll_many_L_L}(b).  Despite the significant
improvement obtained using the fiber taper, we see that the largest
$\xi$ measured is roughly 10 nW$/{\mu}$W, which means that only
1$\%$ of the pump photons are converted to a collected signal
photon, and we should thus consider why $\xi$ is far below unity.  We first note that when considering collection into both directions of the fiber taper, $\xi$ is actually $2\%$ for the standing wave WGMs of the microdisks studied here.  A measure of the fiber taper collection efficiency of the microdisk WGM laser light, $\eta_{0}$, from the passive wavelength scans described above indicate that the external fiber taper collection efficiency should be as high as $\sim22\%$ (corresponding to $11\%$ for the forward transmission direction only).  The total loss through the fiber taper and all of the fusion-splices and connections in the fiber path was measured to be $\sim1.6$dB in these experiments.  If symmetric
loss in the taper about the microdisk coupling region is assumed, this corresponds to a loss of $\sim17\%$ of the WGM laser photons collected
by the taper before they reach the OSA. Taken together, these two factors put an estimate of the upper bound on the fiber-coupled external laser efficiency of $18\%$ for collection into both directions of the fiber.

The roughly order of magnitude difference between the measured ($2\%$) and expected ($18\%$) differential laser efficiency may be a result of several factors involving the complex dynamics within the DWELL active region.  Previous measurements of DWELL injection lasers in
stripe geometries\cite{ref:Liu_G} indicate that the internal quantum efficiency of the quantum dots
is $\eta_{i}^{QD} \sim 0.5$ (this is roughly the percentage of carriers captured by the QDs in the DWELL structure that contribute to stimulated emission above threshold).  This factor can certainly change from growth to growth, and given that the laser threshold 
values are roughly 2-2.5 times higher than that measured in previous work on identically fabricated devices from a different wafer growth\cite{ref:Srinivasan9}, we might qualitatively expect $\xi$ to be reduced by a factor in the range of $2$-$5$ due to $\eta_{i}^{QD}$.

\begin{figure}
\begin{center}
\epsfig{figure=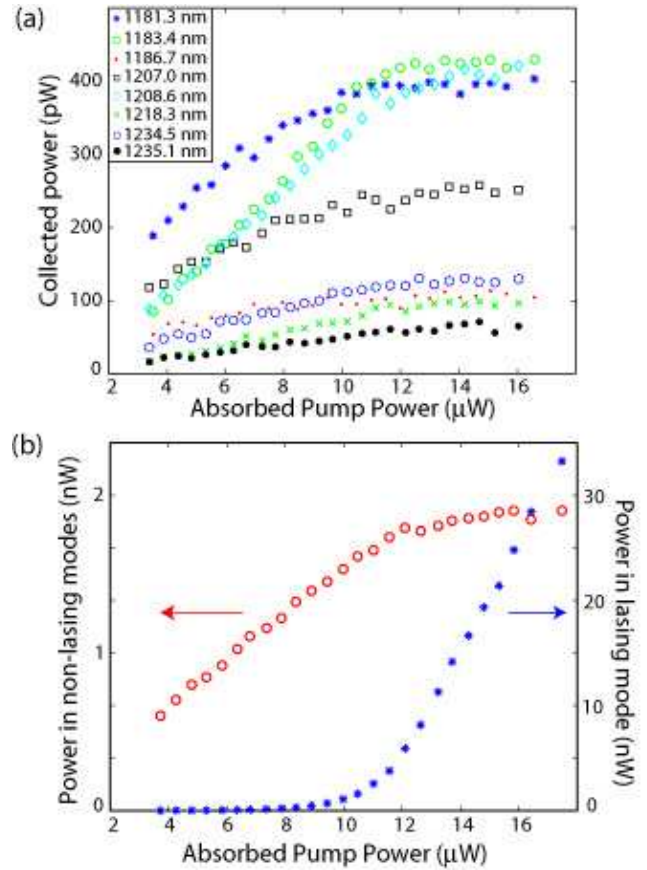, width=\linewidth}
\caption{(a) L-L curves for non-lasing modes of the disk studied in Fig.
\ref{fig:fs_pump_fs_vs_fiber_coll}. (b) Total power in the non-lasing modes (shown by open circles),
showing saturation for pump values close to the threshold value for the lasing mode (shown by asterisks). 
Note that the y-axis scale for the lasing mode is 15 times larger than that for the non-lasing modes.}
\label{fig:saturation_data}
\end{center}
\end{figure}

Both the spectral and spatial distribution of carriers within the microdisk may also lead to reductions in the laser differential efficiency through incomplete clamping of the spontaneous emission into the non-lasing 
modes of the microdisk above threshold.  To
examine such effects in our structures, we measure L-L curves (Fig. \ref{fig:saturation_data}(a))
for a number of the most prominent non-lasing WGMs of the microdisk studied in Fig. 
\ref{fig:fs_pump_fs_vs_fiber_coll}.  We see that the emission into these 
modes is largely clamped above the threshold for the lasing mode (estimated 
to be $10.7 \mu$W of absorbed pump power). The aggregate effect is clearly seen 
in Fig. \ref{fig:saturation_data}(b), where the power into the non-lasing 
WGMs has been summed and plotted along with the L-L curve for the lasing mode.  
Such clamping has been reported by other authors for similarly sized 
microdisks\cite{ref:Slusher2}, while smaller microcavity devices with a larger laser mode spontaneous emission rate, have exhibited a gradual rollover and/or incomplete 
clamping of spontaneous emission\cite{ref:Slusher1,ref:Slusher2,ref:Fujita,ref:Srinivasan4}.  Measurement of the background spontaneous emission into non-WGM, radiation modes of the microdisk was performed using free-space collection (the fiber taper is much more sensitive to WGM emission than to emission from the center of the microdisk into radiation modes), and did show incomplete clamping of the spontaneous emission.  This sort of spatial hole burning has been predicted in numerical modeling of
microdisk cavities\cite{ref:Baba_PECS}.  If this is the case, the effective pump area is limited to a region about the WGM.  Assuming that the WGM radial width is approximately $(\lambda/n_{\text{eff}})$, where $n_{\text{eff}}$ is the effective refractive 
index in the plane of the microdisk\footnote{Finite-element simulations have shown this to be an accurate estimate.}, this corresponds to a $7{\mu}{\text{m}}^2$ area in the devices under test here.  Since the total disk area is $\sim 16 \mu$m$^2$, then only $7/16$ of the pump photons would be effectively pumping the WGM.  
Including this factor brings the expected value of $\xi$ within the range of experimentally measured values.

Aside from reducing taper loss (loss $<0.5$dB can be easily achieved in our 
lab), $\eta_{0}$ is the main parameter that can be improved
upon to increase $\xi$.  This can be done by adjusting the geometry 
of the disk (using thinner disks, for example) to bring the effective
index of the WGMs of the semiconductor microdisks closer to that of the silica fiber taper, so that 
more efficient coupling can be obtained.  A study of such modifications in
Si microdisk structures has been undertaken, and the regimes 
of critical coupling and overcoupling have been achieved\cite{ref:Borselli_coupling}.
In addition, if spatial hole burning is significant, another factor 
that could potentially be improved is the method of pumping.  In particular,
the pumping beam could be shaped to preferentially pump the perimeter 
of the microdisk (i.e., an annular-shaped beam could be used).  Alternately,
as discussed below, a fiber taper could be used to pump the microdisk.

\section{Fiber-pumped microdisk lasers}
\label{sec:fiber_pump}

In addition to improving the collection efficiency, optical fiber
tapers have the potential for improving the pump efficiency of
these QD-containing microdisks; such an effect has in fact been 
demonstrated in previous work on doped glass microcavities\cite{ref:Sandoghdar1,ref:Cai2,ref:Yang_L}.  In particular, if the pump laser is resonant with a WGM of the microdisk, light can be absorbed with high efficiency, and in the case of critical coupling,
complete power transfer can be achieved. This should be contrasted
with the case of free-space pumping, where only a small percent
($10\%$ for the devices we've considered here) of the incident
pump light is absorbed by the device, and some of this absorption
is in a region (the center of the microdisk) that does not
contribute to useful gain for the resonant WGMs\cite{ref:Baba_PECS}. On 
the other hand, fiber taper pumping delivers pump light directly to the perimeter of the
disk.  As we shall discuss in the following section,
the ability to efficiently optically access microcavities is also of particular
importance to future quantum optics experiments.

\begin{figure}[ht]
\begin{center}
\epsfig{figure=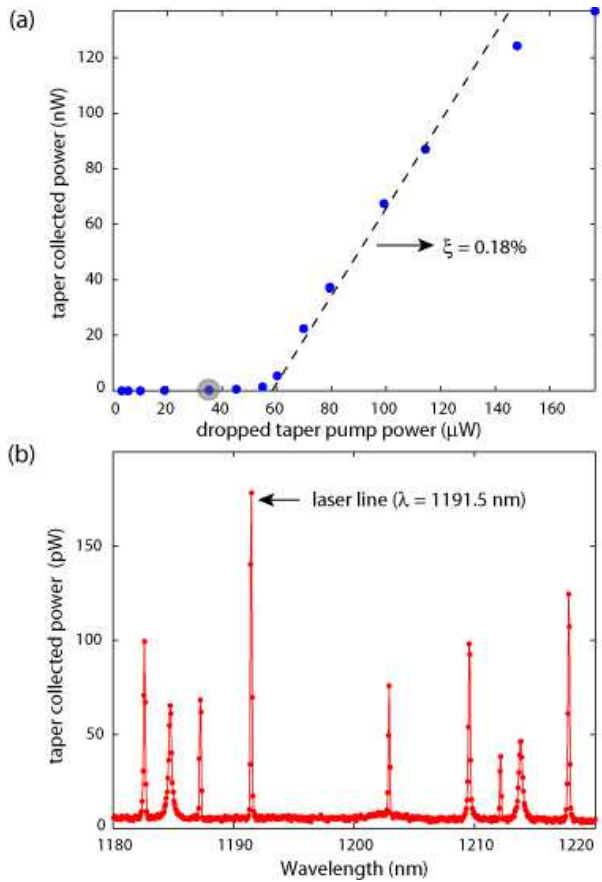, width=\linewidth}
\caption{(a) L-L curve for a QD-microdisk device where the fiber
taper is used for both pumping (at $\lambda=967.6$ nm) and
collection. The absorbed power was estimated to be $66\%$ of the input power in the fiber taper.  (b) Sub-threshold spectrum for this device taken at an estimated absorbed power of $37$ $\mu$W (highlighted point in (a)).}
\label{fig:fiber_pump_data}
\end{center}
\end{figure}

For an initial demonstration, we use a tunable laser operating in
the 980 nm band as a pump source. The 830 nm pump laser is not
used in this experiment because the absorption within the microdisk
at this wavelength is too large to allow uniform pumping of the microdisk perimeter (the pump light is absorbed before a single round trip around the cavity
can be made).  At 980 nm, the material absorption is still relatively high (in particular, the quantum well layer will
be highly absorbing), so that the $Q_{980}$ of WGM modes near the pump
wavelength are not expected to exceed a few hundred.  The 980 nm
tunable laser is connected to the fiber taper input, and the fiber
taper output is connected to the OSA. The taper is contacted to
the side of the microdisk, and the pump wavelength and polarization of
the tunable laser are manually adjusted until the collected power
in the OSA is maximized (this is necessary in order to resonantly
couple to a mode within the pump wavelength band).  A typical L-L curve and sub-threshold spectrum are shown in Fig.
\ref{fig:fiber_pump_data}.  We see that a significant amount of
power is collected into the fiber taper, and that in particular,
the sub-threshold spectrum shows a number of well-resolved modes
with a good signal-to-noise ratio. The estimated absorbed pump power in the microdisk displayed in Fig.
\ref{fig:fiber_pump_data}(a) corresponds to $66\%$ of the input power in the fiber taper,  and is found by taking the difference in the 980 nm band taper
transmission between when the taper is displaced tens of microns above the microdisk (no coupling) and when it is in contact with the microdisk (strongly coupled).  We note that the pump threshold value in this pumping geometry is only about a factor of two less than the
$\emph{incident}$ pump power in the 830 nm free-space pumping, and is significantly larger than what might be expected (ideally, the pump power
here should be less than the $\emph{absorbed}$ pump power in the 830 nm
pumping).  This is most likely a result of the relatively crude method we
have employed to estimate the power absorbed in the microdisk; a much more accurate
method for determining the coupled pump power uses the
wavelength-dependent transmission of the fiber taper to map out the
resonance line due to the WGM at the pump wavelength.  Here, the strong absorption of the microdisk in the 980 nm band makes it difficult to separate resonantly coupled power from scattering losses at the taper-microdisk junction.
%For a WGM in the 980 nm band, we can expect the fraction of the fiber input power resonantly coupled into the microdisk to be roughly $Q_{980}/Q_{1200} \sim 10^{-3}$ times the typical coupling depth in the 1200 nm wavelength band.  As presented above, the coupling depth of WGMs in the 1200 nm band for the microdisk geometry of the devices studied in this work peaks at about $50\%$ with the taper in direct contact with the microdisk, yielding a more realistic absorbed pump power estimate of only $0.05\%$ times the fiber input power.  This corresponds to a laser threshold absorbed power for this fiber pumping geometry of only $50$ nW.  
In order to more carefully study the efficiency of this fiber-pumping and fiber-collecting configuration, experiments in which an excited state of the quantum dots is resonantly pumped through the fiber taper are currently underway.

\section{Discussion and future applications}
\label{sec:discussion}

As mentioned in the introduction of this paper, efficient optical access to wavelength-scale microcavities is of great importance to
quantum optics and information processing applications currently being investigated within cavity QED systems.  In almost any application involving the coherent transfer or manipulation of quantum information, loss is a significant detriment.  As described in Ref. \onlinecite{ref:Kiraz}, current implementations of linear optics quantum computing require a near-unity
collection efficiency of emitted photons from a single photon
source.  The same is true for applications involving quantum repeaters in a quantum network\cite{ref:Cirac}.  A solution that is often proposed is to
embed the single photon emitter within a microcavity with a high
spontaneous emission coupling factor $\beta$, so that the majority
of emitted photons are coupled into the microcavity mode. However,
it is important to note that even for a ${\beta}=1$ microcavity, it is still necessary to have a method to effectively
collect all of the photons that are radiated by that one cavity
mode\cite{ref:Gerard3}.  Also, from a very practical perspective, efficient collection of emitted light from a
microcavity is of premium importance for optical telecommunication wavelengths $>1{\mu}m$, where the dark count rates from single photon counters are
often 2-3 orders of magnitude larger than the Si single photon
counters used at shorter wavelengths\cite{ref:Ribordy}.
 
An efficient coupling channel can also enable a number of
different types of experiments. Having access to this coupling channel makes the
cavity transmission (and reflection) an experimentally accessible
parameter whose behavior can be monitored to detect signatures of
specific types of system behavior.  In recent experimental measurements of coupling between a single quantum dot and a resonant mode of a semiconductor microcavity\cite{ref:Yoshie3,ref:Badolato,ref:Reithmaier,ref:Solomon,ref:Gerard1},
spontaneous emission from the coupled system is the only parameter measured. Alternatively, vacuum-Rabi splitting in a system consisting of a single
QD coupled to a cavity mode can be detected by simply measuring
the transmission through the cavity as a function of the input
wavelength to the cavity; such an experiment is directly analogous
to the experiments done with cooled alkali atoms coupled to a
Fabry-Perot cavity\cite{ref:Thompson,ref:Boca}.  Non-linear effects, such as optical bistability and photon blockade, and coherent control of the quantum system can also be more easily performed through the optical transmission or reflection channel of a microcavity. Perhaps most importantly, by knowing the precise level
of coupling between the fiber taper and the microcavity, the
number of photons injected into the cavity can be precisely
calibrated. This is obviously of paramount importance in
experiments that involve few or single cavity photons.

Finally, it is important to note that while some of the
advantages described above are also available to in-plane
waveguides that are microfabricated next to the microcavities, 
the fiber taper provides a level of flexibility that the
microfabricated waveguides do not, as described in detail in 
Ref. \onlinecite{ref:Srinivasan10}.  In particular, it can
accommodate a number of different cavity geometries, which is very
important for resonators such as microdisk cavities, 
where developing a suitable in-plane waveguide is not
necessarily trivial (in this case, due to the undercut nature of
the microdisks).  Furthermore, fiber tapers can be routinely
fabricated with low losses ($\sim 0.1$-$0.5$ dB in our lab), while
in-plane waveguides still require a device to provide low-loss
coupling into and out of them which poses a similar, albeit slightly less difficult, problem as coupling directly into the microcavity.  In addition, employing 
optical fibers to couple to the microcavities makes these devices easy to integrate
into larger, fiber-optic based networks.  As optical fibers are the low-loss transmission medium of choice, such a fiber-based device architecture would be suitable for implementation into
quantum networks such as those proposed in Ref. \onlinecite{ref:Cirac}.

In conclusion, we have demonstrated the utility of fiber
tapers in the photoluminescence measurements of QD-containing
microdisk cavities.  Improvements in collection efficiency of $1$-$2$ 
orders of magnitude in comparison to normal incidence free-space measurements are
exhibited, and initial measurements on microdisk lasers employing
fiber taper pumping and collection have been presented.  The largest differential laser efficiency was measured to be as high as $2\%$ when accounting for collected light into both directions of the fiber taper.  From direct passive measurements of the microdisks at wavelengths $30$ nm red detuned from the ground-state QD emission peak, the fiber taper collection efficiency of the laser light (including all optical losses of the optical fiber and fiber taper) is estimated to be $18\%$ (again including both directions of emission within the fiber taper).  With further modifications of the microdisk structure to enable better phase-matching of the fiber taper waveguide and the whispering-gallery modes of the high-refractive index microdisk\cite{ref:Borselli_coupling}, significant increases in coupling efficiency, approaching unity, are believed possible with this technique.

\section{Acknowledgements}
The authors thank Matt Borselli, Paul Barclay and Tom Johnson for valuable discussions and assistance in setting up some of the measurement equipment, and Matt Borselli for numerical simulations of the microdisk modes.  K.S. thanks the Hertz Foundation for its graduate fellowship support.

\bibliography{./PBG_6_8_2005}

%\printtables
%\printfigures

%\end{spacing}
\end{document}